\documentclass[preprint]{aastex63}  % preprint is 12pt
\usepackage[utf8]{inputenc}
\usepackage{natbib}
\usepackage{hyperref}
\usepackage{graphics}

\usepackage[width=6.5in,height=9.0in,top=1.8in,left=1.3in,letterpaper]{geometry}
\date{\today}

\NewPageAfterKeywords
\begin{document}
\setcounter{page}{0}

\title{The Scientific Impact of the\\ Vera C. Rubin Observatory's Legacy Survey of Space and Time (LSST)\\ for Solar System Science}

\author{Vera C. Rubin Observatory LSST Solar System Science Collaboration}
\affiliation{Vera C. Rubin LSST Science Collaborations}

\author[0000-0001-5916-0031]{R.\ Lynne Jones}
\altaffiliation{Corresponding author email: lynnej@uw.edu}
\affiliation{DIRAC Institute and Department of Astronomy, University of Washington, Seattle, Washington, USA}
\affiliation{Vera C.\ Rubin Observatory, Tucson, Arizona, USA}

\author[0000-0002-6313-5960]{Michelle T.\ Bannister}
\affiliation{School of Physical and Chemical Sciences | Te Kura Mat$\bar{u}$, University of Canterbury, New Zealand}

\author[0000-0002-4950-6323]{Bryce T.\ Bolin}
\affiliation{IPAC, California Institute of Technology, Pasadena, CA, USA}

\author[0000-0001-7335-1715]{Colin Orion Chandler}
\affiliation{Department of Astronomy \& Planetary Science, Northern Arizona University, Flagstaff, AZ, USA}

\author[0000-0003-3240-6497]{Steven R.\ Chesley}
\affiliation{Jet Propulsion Laboratory, California Institute of Technology, Pasadena, CA, USA}

\author[0000-0002-1398-6302]{Siegfried\ Eggl}
\affiliation{DIRAC Institute and Department of Astronomy, University of Washington, Seattle, Washington, USA}
\affiliation{Vera C.\ Rubin Observatory, Tucson, Arizona, USA}

\author[0000-0002-4439-1539]{Sarah Greenstreet}
\affiliation{DIRAC Institute and Department of Astronomy, University of Washington, Seattle, Washington, USA}

\author[0000-0003-0437-3296]{Timothy R.\ Holt}
\affiliation{Centre for Astrophysics, University of Southern Queensland, Queensland, Australia}
\affiliation{Department of Space Studies, Southwest Research Institute, Boulder, Colorado, USA}

\author[0000-0001-7225-9271]{Henry H.\ Hsieh}
\affiliation{Planetary Science Institute, Tucson, Arizona, USA}
\affiliation{Academia Sinica Institute of Astronomy and Astrophysics, Taipei, Taiwan}

\author[0000-0001-5250-2633]{\v{Z}eljko~Ivezi\'{c}}
\affiliation{DIRAC Institute and Department of Astronomy, University of Washington, Seattle, Washington, USA}
\affiliation{Vera C.\ Rubin Observatory, Tucson, Arizona, USA}

\author[0000-0003-1996-9252]{Mario Juri\'{c}}
\affiliation{DIRAC Institute and Department of Astronomy, University of Washington, Seattle, Washington, USA}
\affiliation{Vera C.\ Rubin Observatory, Tucson, Arizona, USA}

\author[0000-0002-6702-7676]{Michael S.\ P.\ Kelley}
\affiliation{University of Maryland, College Park, Maryland, USA}

\author[0000-0003-2781-6897]{Matthew M.\ Knight}
\affiliation{United States Naval Academy, Annapolis, Maryland, USA}
\affiliation{University of Maryland, College Park, Maryland, USA}

\author[0000-0002-1226-3305]{Renu Malhotra}
\affiliation{Lunar and Planetary Laboratory, The University of Arizona, Tucson, USA}

\author[0000-0001-5750-4953]{William J.\ Oldroyd}
\affiliation{Department of Astronomy \& Planetary Science, Northern Arizona University, Flagstaff, AZ, USA}

\author[0000-0001-5678-5044]{Gal Sarid}
\affiliation{SETI Institute, Mountain View, CA, US}

\author[0000-0003-4365-1455]{Megan E.\ Schwamb}
\affiliation{Queen's University Belfast, Belfast, UK}

\author[0000-0001-9328-2905]{Colin Snodgrass}
\affiliation{Institute for Astronomy, University of Edinburgh, Royal Observatory, Edinburgh, UK}

\author[0000-0002-1701-8974]{Michael Solontoi}
\affiliation{Monmouth College, Monmouth, Illinois, USA}

\author[0000-0003-4580-3790]{David E.\ Trilling}
\affiliation{Department of Astronomy \& Planetary Science, Northern Arizona University, Flagstaff, AZ, USA}
\affiliation{Lowell Observatory, Flagstaff, AZ, USA}

\pagestyle{empty}

% Intro starts at top of page 1
\clearpage
\pagestyle{plain}
\section{Introduction}

Vera C.\ Rubin Observatory will be a key facility for small body science in planetary astronomy over the next decade. It will carry out the Legacy Survey of Space and Time (LSST), observing the sky repeatedly 
in $u$, $g$, $r$, $i$, $z$, and $y$ over the course of ten years using a 6.5~m effective diameter telescope with a 9.6 square degree field of view, reaching approximately $r=\,$24.5 mag (5-$\sigma$ depth) per visit.
The resulting dataset will provide extraordinary opportunities for both discovery and characterization of large numbers (10--100 times more than currently known) of small solar system bodies, furthering studies of planetary formation and evolution.
%The resulting 10--100 factor increase in known population sizes will further studies of planetary formation and evolution as they occur in our own solar system. 
 This white paper summarizes some of the expected science from the ten years of LSST, and emphasizes that the planetary astronomy community should remain invested in the path of Rubin Observatory once the LSST is complete.
\looseness=-1

%%
%\begin{figure}
%\begin{minipage}{.5\textwidth}
%    %\centering
%    \includegraphics[width=0.9\textwidth]{InnerSSAvsIblk.png}
%    \caption{Inner Solar System showing small body populations in inclination (Inc) and semi-major axis: near-Earth asteroids in green, with the subpopulations as follows: Atiras in aquamarine; Atens in chartreuse, Apollos in sea green and Amors in dark green. Main-Belt Asteroids are shown in blue, the Hildas in red, Jovian Trojans in purple, and the Jupiter family comets in olive. The inner planets (Mercury, Venus, Earth, Mars \& Jupiter - orange) are also indicated. After \citet{Horner2020}.}
%    \label{fig:innerSS}
%\end{minipage}
%\begin{minipage}{.1\textwidth}
%\end{minipage}
%\begin{minipage}{.5\textwidth}
%    %\centering
%    \includegraphics[width=0.9\textwidth]{OuterSSAvsIblk.png}
%    \caption{Outer Solar System showing Small Body populations with %the Inner sysetem colors in fig. \ref{fig:innerSS}.  Centaurs (brown) are shown between Jupiter and Neptune. The Neptune Trojans (orange-red) can be seen at 30 au, and the Plutinos (deep pink) at 39.5 au, just interior to the TransNeptunian Objects (orchid, between ~40 and 48 au). To higher eccentricities, the Scattered Disc objects (maroon) can be seen spreading outward in a curve. Two cometary populations are shown, the Jupiter family comets (olive) and the Halley type comets (cyan). After \citet{Horner2020}.}
%    \label{fig:outerSS}
%\end{minipage}
%\end{figure}

\section{The LSST as a small-body discovery machine}

%\assignedto{Mario, Lynne}

Discovery and census of objects in the solar system is one of the four foundational science cases for the LSST \citep{2019ApJ...873..111I}.
The suitability of the final LSST cadence for the discovery and characterization of solar system objects will therefore be a high-priority consideration in its selection.
%The LSST cadence will therefore be selected so as to include elements to enable the discovery and characterization of solar system objects using well understood moving object linking algorithms.
A solar system object observed by the LSST will be identified as such with 95\% efficiency (on average) if it was detected on at least three nights within a window of 15 days, with a minimum of two visits per night \citep{2019ApJ...873..111I}. 
Realistic simulations performed using the Moving Object Processing System (MOPS; which was developed by the Pan-STARRS survey and adopted by LSST) show $>$99\% linking efficiency across all classes of solar system objects \citep{2013PASP..125..357D}, and at least 93\% efficiency for Near-Earth objects \citep[NEOs;][]{2017AJ....154...12V,2017AJ....154...13V} using this criterion.
%The same criterion has been used in NASA studies and is confirmed as reliable by a detailed analysis of orbital linking and orbit determination using Moving Object Processing System (MOPS) code \citep{2005AAS...20712102J,2017AJ....154...12V,2017AJ....154...13V} developed by the Pan-STARRS survey (and adopted by LSST). % in a collaborative effort with Pan-STARRS).
%The MOPS software system and its algorithms are significantly more advanced\footnote{The baseline algorithm has recently been upgraded to HelioLINC \citep{2018AJ....156..135H}} than anything previously fielded for this purpose to date. 
%Realistic MOPS simulations show $>$99\% linking efficiency across all classes of solar system objects \citep{2013PASP..125..357D}, and at least 93\% efficiency for near-Earth objects \citep[NEOs;][]{2017AJ....154...12V,2017AJ....154...13V}.
Newly discovered objects will be reported to the Minor Planet Center, generally within 24 hours of discovery.

The LSST alert stream will
%is a crucial component that inherently supports solar system science. The alert stream
identify all transient objects in each image, including any small body detections or trailed objects, and distribute alerts within 60 seconds of shutter closure.
%All known solar system objects, including comets and fast moving (streaked) objects, will be identified as such in the alert packet.
%This continuous stream of data will allow for rapid detection of objects with unexpected changes in brightness or morphology.
%, as well as rapid detection and follow-up of trailed (nearby) objects.
In total, the LSST will obtain approximately 1.8 billion observations of 5.5 million objects %(over 80\% of which will be discovered by LSST itself) 
over 10 years. Data will be automatically reduced and delivered with systematic-limited astrometric and photometric precisions of 10 mas and 0.01 mag, respectively. Rubin Observatory will be by far the most prolific discovery and characterization machine of small solar system bodies  throughout the 2020s. 
%all populations of small bodies of the Solar System in 2020s.
\looseness=-1

\begin{figure}[htb!]
    \centering
    \vspace{-10pt}
    \includegraphics[width=0.8\linewidth]{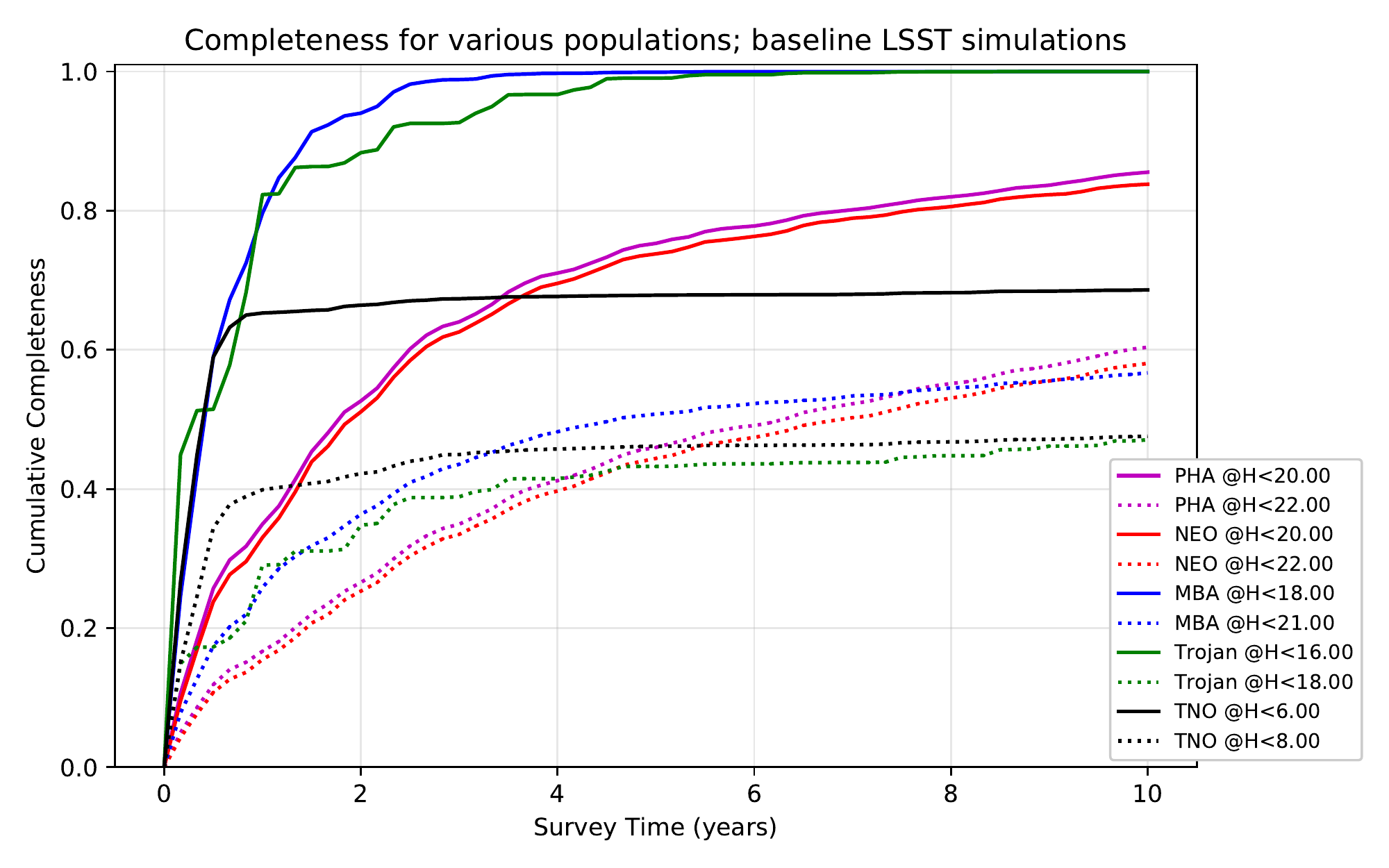}
    \vspace{-10pt}
    \caption{
    LSST discovery completeness as a function of time, for various modeled solar system populations \citep{granvik2018debiased,grav_ssm,2011AJ....142..131P,2009AJ....137.4917K}. Cumulative completeness is reported for two different absolute magnitude ($H$) values; one near the bright end of the population (solid lines) and one near the 50\% completeness level (dashed lines). These follow similar tracks over time, showing that most objects are discovered early in the survey.
    %Discovery completeness as a function of time, for various Solar System populations as detected by LSST. The PHA and NEO population is based on \citet{granvik2018debiased}, the main-belt asteroid and Jovian Trojan population is based on \citet{grav_ssm}, and the TNO population is based on the CFEPS L7 model \citep{2011AJ....142..131P, 2009AJ....137.4917K}. The cumulative completeness is reported for two different absolute magnitude ($H$) values; one near the bright end of the population (solid lines) and one near the 50\% completeness level (dashed lines). Generally these follow a similar track over time, showing that most objects are discovered in the first few years of the survey. The long synodic periods of NEAs slows the discovery rate of these objects; for comets and interstellar objects, it is likely that a similar slowly increasing rate of discovery over time would be present, due to their significantly more eccentric orbits.
    } 
    \label{fig:completeness_time}
\end{figure}

Table~\ref{tab:discoveries} summarizes estimated LSST discovery yields for several major solar system populations. Most discoveries will occur during the first few years of the survey (Figure~\ref{fig:completeness_time}). %Significant fractions of these populations 
A significant fraction of these objects
will receive multi-band observations suitable for measuring colors and lightcurves.
Beyond these major populations, the LSST will also discover an order of magnitude more interstellar objects, active asteroids, Centaurs, planetary Trojans, 
%irregular satellites, 
temporarily captured objects (i.e., ``mini-moons''), Hildas, and inner Oort Cloud objects.
%as well as subgroups of these populations, such as Hildas or inner Oort Cloud Objects. 
\looseness=-1

\begin{table}
    \centering
    \caption{Small-body population numbers as of today (7/2020; JPL Small-Body Database) and after LSST (approximate predicted LSST results based on simulations).}
    \label{tab:discoveries}
    \vspace{-10pt}
    \begin{tabular}{lcccc}
        \hline\hline
        Population & Currently known (approximate) & LSST discoveries (predicted) \\
        Near-Earth objects & 23\,000 & 100\,000\\
        Main Belt asteroids & 	856\,000 & 5\,000\,000\\
        Jovian Trojans & 8\,000 & 280\,000 \\
        Trans-Neptunian objects & 3\,500 & 40\,000 \\
        Comets & 4\,000 & 10\,000 \\
        Interstellar objects & 2 & $>10$ \\ \hline
    \end{tabular}
\end{table}

\section{Science enabled by LSST}

%\note{about 2 to 3 paragraphs each, or 4 to 5 pages total}

\subsection{Near-Earth objects}
%\assignedto{Siegfried}
The LSST plays a crucial role in achieving the congressional target of discovering 90\% of all potentially hazardous asteroids (PHAs) with $H\approx22$ \citep{2017AJ....154...12V,jones2018large}. Current predictions suggest that the LSST alone could be responsible for a five-fold increase in the total number of known NEOs by the end of the survey (Table~\ref{tab:discoveries}), detecting on the order of 100\,000 NEOs at a range of sizes down to $H<25$~mag. %($\approx65\%$ of all PHAs greater than 140\,m).
%on potentially hazardous orbits.

The LSST's excellent astrometric precision ($\sigma\approx10$\,mas) is expected to facilitate detections of non-gravitational forces acting on orbits, including the Yarkovsky effect, solar radiation pressure, outgassing, and collisions. Understanding how these effects influence NEO trajectories is critical to our ability to precisely predict future impacts \citep{farnocchia2013yarkovsky}.

The timely processing and publication of NEO observations allows impact monitoring services to provide advance notice of close approaches and potential impacts. This facilitates critical characterization efforts including radar, spectroscopic, and light curve observations.
Variations in the brightness of NEOs can also yield clues on NEO disruption mechanisms at small perihelion distances \citep{granvik2016super}, which can then be utilized to probe NEO internal structure and test dynamical models, both critical for effective planetary defense.

Yearly releases of NEO catalogs exclusively based on LSST observations enable unbiased population predictions of unprecedented quality. Unbiased populations can be used to quantify the long-term impact flux of NEOs as a function of size \citep{granvik2018debiased}. The orbital, absolute magnitude, and taxonomy distributions within the NEO population derived from the LSST data can help identify correlations between taxonomy and orbital properties for NEOs, as well as determine the orbital distribution of objects down to ten meters in size.

\looseness=-1

\subsection{Main Belt asteroids}
%\assignedto{Tim, Henry}

Main Belt asteroids (MBAs)
%and other small inner solar system bodies such as the Hildas and Jovian Trojans 
represent important compositional and dynamical tracers of the solar system's formation and evolution. Their orbital element distribution, size-frequency distributions, and total mass provide key constraints for planetary migration models like the Nice Model and the Grand Tack Model \citep{morbidelli2010_nicemodel,walsh2011_grandtack,yoshida2019_sfds}, 
while taxonomic classifications %of large numbers of main-belt asteroids
provide additional constraints on those models and also enable the tracing of more recent dynamical and physical evolution \citep[e.g.,][]{demeo2014_astbeltmapping}. 
The LSST will discover an order of magnitude more MBAs than are currently known,
%permitting improved characterizations of inner solar system small body populations,
%modeling of the distribution of asteroids near the edges of resonances, 
providing new insights into solar system formation and evolution.
%LSST's high discovery rate will also facilitate the discoveries of rare objects, such as retrograde asteroids and active asteroids (Section~\ref{subsec:spcs_actvasts}).
\looseness=-1

%For illustration, LSST will obtain about 200--300 photometric measurements per object for over 5 million main-belt asteroids during its 10-year survey (see Section 5 in \citealt{2009arXiv0912.0201L}).  
%, or no larger than 0.05 mag even in the most pessimistic case of a faint asteroid that is always at the LSST detection limit. Based on past analyses of SDSS asteroid data, it is clear that with color uncertainties in this range, the LSST dataset will be revolutionary.
% rework these two paragraphs to focus on asteroid families & how LSST supports this?
%Detailed compositional information, such as results from the Sloan Digital Sky Survey, has allowed for the confirmation of the compositional homogeneity of families \citep{ivezic2002_astfamilies}  -- delete? or add to below? 
Beyond discovery, the multi-band, multi-epoch observations from the LSST \citep[$\sim$200--300 photometric measurements per object for over 5 million MBAs during its 10-year survey; Section 5 in][]{2009arXiv0912.0201L} will enable detailed physical studies.
Precise multi-band optical colors will enable robust and efficient taxonomic
classification of MBAs \citep{2010A&A...510A..43C}, and detailed studies of asteroid families \citep[e.g.,][]{ivezic2002_astfamilies,MOC2008}, where
%LSST multi-band measurements will not be simultaneous. Nevertheless, 
%given the small color variability of asteroids \citep{2004MNRAS.348..987S}, the rotational period and light curve can be estimated by combining all bands using modern approaches to multi-band periodograms \citep{2015ApJ...812...18V}. 
%in cases where good light curve fits \citep[e.g.,][]{2015ApJ...812...18V} can be found,
%the resulting
color precisions on the order of 0.01 mag for sufficiently bright asteroids are expected to be achievable by the LSST.
When combined with infrared data, optical albedos can be estimated
and used to improve taxonomic and family classifications \citep{2012ApJ...745....7M}, and
to derive physical size estimates \citep{MMI, IvezicIvezic2020}. 
\looseness=-1

\subsection{Short-period comets and active asteroids}
\label{subsec:spcs_actvasts}
%\assignedto{Matthew, Colin, Henry}
% phrases we need: JFC, MBC

Active objects are small solar system bodies that exhibit any type of mass loss, whether driven by ice sublimation, impact disruption, rotational destabilization, or other mechanisms. 
These bodies inform us about the distribution of volatiles throughout the solar system and hold clues about solar system formation. 
%They are found everywhere from the near-Sun environment (Sun-grazers) to the Oort cloud and beyond 
They include Jupiter-family comets (JFCs), Oort Cloud comets (Section~\ref{subsec:oortcloudcomets}), active asteroids \citep{jewittActiveAsteroids2015a,snodgrass2017_mbcs}, active Centaurs \citep{jewitt2009_actvcentaurs}, and active interstellar objects (Section~\ref{subsec:interstellar}).  %In this section, we focus on active objects in the inner solar system, namely JFCs and active asteroids.
%, which in turn consist of main-belt comets \citep[MBCs;][]{hsieh2006_mbcs,snodgrass2017_mbcs}, whose activity is sublimation-driven, and disrupted asteroids, whose activity is related to other processes like impacts or rotational destabilization \citep{jewitt2011_scheila,chandlerSixYearsSustained2019a}.
\looseness=-1

An order of magnitude increase in the known JFC population from the LSST
%$\sim$600 currently known), 
%as LSST is expected to produce, 
will facilitate improved modeling of their source regions and evolutionary processes \citep{brasser2015_jfcpopulation} and enable large-scale photometric and spectroscopic follow-up studies that will improve our ability to identify and characterize distinct taxonomic classes \citep{cochran2015_comettaxonomy}.  Meanwhile, only about 30 active asteroids are currently known, making them difficult to characterize on a population-scale basis, given the diversity exhibited by the population thus far \citep{jewittActiveAsteroids2015a}.
%Recent estimates place the rate of active asteroids among the total main-belt population at $\sim$1:10,000 \citep{chandlerSAFARISearchingAsteroids2018b}, where about half of those may be main-belt comets \citep{hsieh2015_ps1mbcs}, but these rates may be higher for LSST given its greater image depths compared to current surveys.  
The identification of many more active asteroids by the LSST will both enable population-level studies, such as efforts to ascertain the rates of rotational and impact disruptions in the asteroid belt \citep{denneau2015_asteroiddisruptions,mcloughlin2015_asteroidcollisions}, and increase the number of individual objects for targeted studies.
\looseness=-1

%At present several groups of active objects have few known members (e.g., $\sim$30 active asteroids, $\sim25$ active Centaurs), thus they are poorly understood. 
%For example, recent estimates place active asteroid occurrence at $\sim$1:10,000 \citep{chandlerSAFARISearchingAsteroids2018b}, about half of which may be main-belt comets \citep{hsieh2015_ps1mbcs}, but these rates are predominantly limited by survey magnitude limits \citep{jewittActiveAsteroids2015a}. Similarly, some activity mechanisms are transient in nature (e.g., impact events), consequently so few have been observed that occurrence rates are yet unknown. [Mention family associations here, reference \cite{hsiehAsteroidFamilyAssociations2018}; can connect to importance of e.g., Gault/Phocaea but maybe too much detail?]

The LSST's 10-year timescale is well-suited for the study of active asteroids and short-period comets, as all active asteroids and about half of all known JFCs have orbital periods of less than 10 years. LSST should produce data sets for a large number of these objects around at least one full orbit, providing at least some of the following: constraints on the initiation/cessation and seasonal variation of activity \citep{AHearn1995}, coarse sampling of outbursts as functions of orbit position
%that will yield insight into the mechanism(s) causing these poorly understood phenomena 
\citep{Ishiguro2016}, nucleus size estimates \citep{fernandez2013_seppcon}, and rotational lightcurves to constrain rotation periods, axis ratios, and pole orientations, and identify binarity or higher-order multiplicity \citep{Lamy2004,Kokotanekova2017}.  The LSST's well-sampled data set will also enable systematic studies of active object nuclei before and after active events, helping us to better understand small body interiors, space weathering, and the effects of cometary outgassing \citep[e.g.,][]{bodewits2014_scheila}.  Extending the initial 10-year survey would permit many comets to be observed over multiple orbits, investigating the secular evolution of cometary activity on an unprecedented scale \citep[e.g.,][]{2019ApJ...886L..16K,Nesvorny2017}. %, yielding new insights into cometary evolution, activity lifetimes, and solar system formation \citep[e.g.,][]{2019ApJ...886L..16K,Nesvorny2017}.
\looseness=-1

\subsection{Trans-Neptunian Objects}
\label{subsec:tnos}
%\assignedto{Lynne, Henry, Colin, Will} % Will Oldroyd/Colin on Planet 9

% COC: we need to define Centaurs I think given the 3+ broad definitions out there
% If it helps, latest active Centaur (2014 OG392) was discovered in DECam archival data, similar survey depth. -- Colin
%active Centaurs \citep{jewitt2009_actvcentaurs}

Trans-Neptunian Objects (TNOs), Centaurs, and Scattered Disk Objects (SDOs) provide key insights into planetary formation and evolution, %since the discovery of the first TNO, 1992 QB$_1$ \citep{1993Natur.362..730J}.
recording the imprint of the formation and migration of the giant planets in their orbital distributions and physical properties. 
%These populations are comprised of significantly dynamically processed objects which have recorded the imprint of the formation and migration of the giant planets in their orbital distributions and physical properties. 
Decoding these imprints has been hindered by the lack of sufficiently large population samples with high precision orbits and well understood discovery circumstances. The LSST will discover on the order of tens of thousands of TNOs, SDOs and Centaurs, and will deliver these objects with extremely precise orbits (due to high-precision astrometry spanning multiple years). Many of the objects will have multi-color photometry suitable for calculating colors and lightcurves for rough composition and rotation rate distributions. 

The discovery, orbital classification, and physical characterization of large numbers of TNOs will help address the origin of the cold classical Kuiper belt, determining if these objects are primordial or were implanted into these orbits early in the history of the solar system, and ascertain its connections with other populations, such as Oort Cloud comets, Centaurs, and planetary Trojans \citep[e.g.,][]{2018ARA&A..56..137N}.
%Dynamical modeling, particularly when coupled with information on physical composition, can evaluate links between these populations and other populations throughout the Solar System, such as Oort Cloud comets, Centaurs, and planetary Trojans
%, building a cohesive model of planetesimal formation and migration 
%\citep[e.g.,][]{2018ARA&A..56..137N}.
%Precise measurement of the distribution of resonant TNOs, especially with the exquisite orbits from LSST, will map the libration islands of high-order resonances, constraining models of the migration of Neptune. 

Due to its depth and wide sky coverage, the LSST will discover many more high perihelia, large semi-major axis TNOs, such as inner Oort cloud objects ({\it i.e.} Sedna-like objects) and ``extreme'' TNOs (with $q>40$~AU and $a>150$-250~AU), than the $<20$ that are currently known.
% Double checking with JPL Horizons (17) and MPC (18): Fewer than 20 ETNOs/IOCs (a>150 & q>40) are known
%Fewer than 50 of these objects are currently known, and their origin is poorly understood. 
A larger sample with well understood discovery circumstances will help ascertain whether their source is related to a hypothesized distant planet \citep{2014Natur.507..471T, 2019PhR...805....1B}, Neptune migration, stellar perturbations, or some other mechanism \citep[e.g.,][]{2020tnss.book...61K}, and also whether their observed orbital alignment is real (possibly indicating the presence of a distant planet) or the result of observational bias \citep[e.g.,][]{2020tnss.book...79T,2019AJ....157...62B,2017AJ....154...50S}. Additionally, the LSST may directly detect a distant planet if it is sufficiently bright and within the survey footprint, otherwise ruling out its presence in 61\% of the sky \citep{2018AJ....155..243T}. %An increased number of known extreme TNOs will also help to confirm whether their observed orbital alignment is real (possibly indicating the presence of a distant planet) or the result of observational bias \citep[e.g.,][]{2020tnss.book...79T,2019AJ....157...62B,2017AJ....154...50S}.
%or show that it is the result of observational bias  \citep{2017AJ....154...50S}.
A potential survey extension using deeper (10~minute, $r\approx26$) exposures would increase the survey sensitivity for more distant and smaller TNOs. 
\looseness=-1

% This paragraph was fun, but could be cut. It's covered in the science book and roadmap, basically.
%The determination of colors, and any correlation with dynamical properties, for large numbers of TNOs will further constrain models of planetary formation. Colors can also be used to help identify collisional families in the outer Solar System, such as Haumea. Determination of rotational light curves, especially if combined with tighter constraints on binary fraction, for a range of dynamical classes, can further constrain planetary evolution models as well as place constraints on planetesimal formation. Finally, accurate and precise astrometry for these distant TNOs can be used to predict stellar occultation opportunities, allowing direct measurements of object sizes. 

%A productive survey extension could involve going deeper to detect more distant or smaller TNOs, using 10~minute exposure times instead of 30~seconds to reach depths of about $r=26$.
%if the standard 30~second exposure time was increased to 10~minutes, 
%the typical visit depth would be about $r=26$. 
%With Rubin Observatory's large field of view, between 200 and 400 square degrees could be observed each night with 10~minute exposures, covering the sky $\pm$10$^{\circ}$ from the ecliptic three times per month over a season. 
%With Rubin Observatory's large field of view, between 200 and 400 square degrees could be observed each night (depending on whether one or two visits per pointing were required); this would still be sufficient to cover the ecliptic $\pm$ 10 degrees three times per month over a season. 
\looseness=-1

\subsection{Oort cloud comets}
\label{subsec:oortcloudcomets}
%\assignedto{Michael, Mike}

As the most distant solar system population, the Oort Cloud and 
%and only weakly held by the gravity of the Sun, 
its structure reflect the early evolution of the solar system and that of the local Galactic environment \citep{1986Icar...65...13H, 2011Icar..215..491K}. Oort Cloud objects are beyond current observational limits in-situ, however, and thus Oort Cloud comets (OCCs) provide our only opportunities to probe the physical and dynamical properties of these objects. %Moreover, these comets are important comparators for the short-period comet population, which, beyond the difference in dynamical origins, has evolved by insolation due to their smaller orbits and frequent perihelion passages (semi-major axes $\lesssim10$~au vs.\ $10^3-10^4$~au). 
The LSST is expected to discover thousands of OCCs \citep{2010Icar..205..605S}, with generous photometric orbital coverage thanks to the survey's exceptional single-visit sensitivity limit.
\looseness=-1

The discovery of distant OCCs, with perihelion distances, $q$, $>10$~au, enables follow-up characterization of a population that has likely never been closer to the Sun than Jupiter, and therefore in an earlier evolutionary phase than OCCs with $q<5$~au.  At this stage, comet orbits are less altered by planetary perturbations and non-gravitational forces, and offer a rare opportunity to place constraints on the Oort cloud population.  Furthermore, the discovery of comets with perihelia beyond 15~au directly tests the existence of the inner Oort cloud \citep{2019AJ....157..181V}.  Currently, no comets have been discovered with $q>12$~au \citep[as of July 2020;][]{1996DPS....28.2504G}, making this an important discovery space for LSST.  %We anticipate ground-breaking follow-up studies of these objects with the James Webb Space Telescope (JWST).

Early detection of inbound OCCs provides us with the opportunity to study the active behavior
%(i.e., mass-loss mechanisms) 
of this population.  A long standing problem with our current knowledge of the OCC population is the apparent paucity of long-period objects, the so-called long-period comet fading problem.
%, originally noted by \citet{oort50}.  
Ad-hoc models of OCC fading (e.g., splitting, disintegration, depletion of volatiles) have succeeded in addressing the missing comets \citep{1999Icar..137...84W}, but physically-motivated explanations
%interpretations of the fading law(s) 
are still needed \citep{2019AJ....157..181V}.  With an increased discovery rate and broad orbital coverage of OCCs, LSST will enable the study of key aspects of OCC evolution, such as a statistical description of when comets begin phases of activity, undergo fragmentation or disintegration, and the follow-up study of these events. 
%The observational cadence and depth will provide the needed observations of such events to allow coordinated observation campaigns from ground and space-based facilities.
\looseness=-1

\vspace{-12pt}
\subsection{Interstellar objects}
\label{subsec:interstellar}
%\assignedto{Matthew}

%Interstellar objects (ISOs), small bodies on hyperbolic orbits that originated outside of our Solar System, are expected as a natural consequence of planet formation and solar system evolution \citep[e.g.,][]{oort50}. 
%\citet{Engelhardt2017} argued that the lack of discoveries to that point implied that our Solar System may be unusual or that our understanding of the processes at work was too simplistic. 
Detections of the first two known interstellar objects, 1I/`Oumuamua in 2017 \citep[e.g.,][]{2017Natur.552..378M} and 2I/Borisov \citep[e.g.,][]{Fitzsimmons2019} in 2019, have given the first glimpses of macroscopic material from other solar systems and revealed that the range of properties of small bodies is broader than previously known.
%\looseness=-1
%Prior to the discovery of 1I, it was predicted that LSST would discover 0.001--10 ISOs during its 10-year lifetime \citep{Cook2016}. 
%Since inactive objects are fainter than cometary objects for a given nucleus size, 
%The discovery of 1I as an apparently asteroidal object constrained the rate to $\sim$1 per year  \citep{Trilling2017}. The discovery of 2I relatively soon thereafter suggests that the frequency may be even higher, but it now seems reasonable to estimate that LSST will discovery of order ten interstellar objects during its first ten years. While each newly discovered ISO will, undoubtedly, be studied in great detail, the collective statistics about interstellar objects will reach critical mass to enable meaningful tests of planet formation mechanisms and planetesimal ejection rates in other star systems.
%[More specific examples: asteroid/comet, size distribution, assess if fragment population, mass lost per star]
Based on the discoveries of 1I and 2I, it now seems reasonable to estimate that the LSST will discover on the order of ten interstellar objects (ISOs) during its first ten years. While each newly discovered ISO will undoubtedly be studied in great detail, collective statistics for the population must reach critical mass to enable meaningful tests of planet formation mechanisms and planetesimal ejection rates in other star systems.
Continued surveying by Rubin Observatory beyond the initial 10 year LSST would be highly beneficial to the study of ISOs by improving population statistics. 
%Though the number of known ISOs after 10 years may be an order of magnitude larger than today, it will still be small enough that significant advances will be gained simply by continuing the survey for a longer time baseline. That said, 
%A future survey optimized for solar system objectives could facilitate even greater advances, e.g., deeper imaging to search more space and reveal fainter objects.
%If small or inactive ISOs are more common than currently thought, such an improvement would disproportionately increase the number of discoveries. 
A future survey extension with deeper imaging
%, maintaining sufficient cadence across the sky, 
would increase the survey volume 
to allow more robust investigation of the ISO size distribution and reveal whether it is reflective of these objects' formation or evolution, insight critically needed in order to understand the context for 1I and 2I \citep[cf.][]{Oumuamua2019}. 
% Matthew says: I acknowledge the following is out there, but our ISSI 'Oumuamua team has a paper in prep discussing this. I have no objections to not including it in the White Paper.
%The longer baseline would also increase the chance of detecting the first {\it intergalactic} object(s) which
%, if detected in sufficient numbers,  could be used as probes of local galactic dynamics.

%\note{[From NEOs]Interstellar objects: broadly consisting of objects on orbits inward of or diffusing inward from the asteroid belt and objects on unbound orbits passing through the Solar System, like `Oumuamua  \cite{2017Natur.552..378M}.} % (Meech et al. 2017)

%\note{[from science roadmap] Discovery/frequency/population estimates of interstellar objects on unbound orbits passing through the Solar
%System as a potential probe of planet formation and planetesimal ejection rates in the local solar neighborhood.}

\section{Connections with Missions and Other Facilities}

%\textit{Lucy} is a NASA Discovery class mission set to launch in 2021 targeting . LSST is expected to discover several orders of magnitude more Jovian Trojans within the first few years, which will offer additional opportunities to extend the \textit{Lucy} mission with potential new targets of opportunity. LSST will also provide context for the \textit{Lucy} mission. 

%The NASA \textit{Psyche} Discovery Mission is planed to visit the potentially metallic world, 16 Psyche. The current hypothesis is that it is the exposed core of a larger object \citep{Hardersen2005Mtypesexposed}. While 16 Psyche, has a visible magnitude that would saturate in LSST, \textit{Psyche} will observe and map the surface of the object, LSST can potentially offer links to the remnant fragments of the collision. This extends the scientific outcomes of the \textit{Psyche} mission and places it within the wider context of the asteroid belt. 

%\subsection{Synergies with other missions}

Previous studies combining SDSS asteroid data with data from other ground and space-based sources forecast strong synergistic value of the LSST dataset (e.g., \citealt{MOC2008}; \citealt{2010A&A...510A..43C}; \citealt{2012ApJ...745....7M}). Both the multi-color nature and well-sampled light curves for an unprecedentedly large sample of small bodies will be exceedingly valuable for other missions. 
%First and foremost, given LSST cadence and the sheer dataset size, discovery of rare and highly valuable objects as candidates for follow-up with other facilities will be much more probable than with existing datasets. For example, LSST could discover several interstellar visitors, such as `Oumuamua, per year. Such rare objects 
Follow-up of rare objects discovered by the LSST will be a high priority for facilities as the James Webb Space Telescope and 30m-class telescopes like the Thirty-Meter Telescope, Giant Magellan Telescope, and Extremely Large Telescope. Deep coordinated searches 
%for very faint moving objects 
with missions such as the Roman Space Telescope would open new ways to probe small and cold solar system objects.  Meanwhile, combining data from the LSST and the upcoming NEO Surveillance Mission spacecraft will allow determination of diameters and albedos for an unprecedented number of solar system objects \citep[cf.][]{mainzer2019_neowisealbedos},
while LSST lightcurves will be essential for assembling data collected over several days by NASA's SPHEREx mission, which is expected to collect 0.75-5.0 $\mu$m spectral data for $\sim$200,000 asteroids, into self-consistent spectra for each object.

%LSST will provide a unique synergy with the NASA's SPHEREx mission.  
%NASA's SPHEREx mission will obtain a minimum of four 0.75-5.0 micron spectra at every point along the ecliptic, achieving point source sensitivities deeper than 2MASS in every spectral element, and will deliver at least four such spectra for about 200,000 asteroids. In the case of asteroids, light curve information will be required to assemble SPHEREx measurements into self-consistent spectra for each object, and as such, LSST lightcurves will be essential for the analysis of SPHEREx asteroid data.

%SPHEREx will observe the entire sky multiple times during its planned two-year mission. Its overlapping scan strategy will obtain a minimum of four 0.75-5.0 micron spectra at every point along the Ecliptic. SPHEREx will achieve point source sensitivities magnitudes deeper than 2MASS in every SPHEREx spectral element, and will deliver at least four such spectra for about 200,000 asteroids. However, in case of asteroids SPHEREx measurements for a given spectrum will be collected over 3 days and the light curve information will be required to assemble measurements into a single self-consistent spectrum. LSST will help greatly by providing the required light curves for essentially all asteroids detectable by SPHEREx.

Discoveries from the LSST will help to enhance the upcoming NASA Discovery missions Lucy (launch date 2021), targeting seven Jovian Trojans,
%including the binary Patroclus-Menoetius system, 
and Psyche (launch date 2022), targeting potentially metallic asteroid (16) Psyche.  
%LSST is expected to discover several orders of magnitude more Jovian Trojans than are currently known within the first few years of the survey, which 
These can offer opportunities to extend the Lucy mission with new targets of opportunity \citep{schwamb2018_lucy}, while observations of larger populations of similar asteroids will add context to the results of the missions. 
% and extend the insights gained from those missions to larger populations of objects that are impossible to study with targeted missions alone.

% Comet Interceptor text from Matthew:
ESA's Comet Interceptor mission \citep{Snodgrass2019} is scheduled to launch in 2028 and wait at the Earth-Sun Lagrange point L2 for up to $\sim$3--5 years until a suitable long-period comet or ISO is identified for a flyby, potentially making the first ever in situ observations of a comet entering the inner solar system for the first time. 
%As the target is not yet known, the mission relies on surveys detecting an accessible object sufficiently early to allow the spacecraft to intercept it. 
The LSST's survey power makes it very likely to be the initial discoverer of the eventual target, and the expected discovery at large heliocentric distance will be critical for mission optimization.

%Comet Interceptor, Lucy, Discovery program

%NEOWISE, NEOSM, Roman Space Telescope, TMT, GMT, EELT

%General context to all missions (past and present)

\section{Conclusions and Proposed Actions}
%\assignedto{Mike, Lynne, (Zeljko)}

%We recommend that the upcoming Planetary Science Decadal Survey include the following findings:
%\begin{itemize}
%    \item 
Rubin Observatory and LSST can be transformative for planetary astronomy. An exciting, but non-comprehensive, list of science impacts has been summarized here; additional impacts are outlined in the LSST Solar System Science Collaboration roadmap \citep{2018arXiv180201783S} and the LSST Science Book \citep{2009arXiv0912.0201L}.  The LSST will significantly increase the discovery rates of new objects in small-body populations from NEOs to TNOs and active objects, and enable large-scale investigations of the physical and dynamical properties of small bodies. In addition, the LSST can support future spacecraft missions by helping to discover primary, flyby, and extended mission targets and providing context to the results of targeted missions.

%    \item 
{\bf Effective use of Rubin Observatory for solar system studies is contingent on preparedness now, as well as strong grant support once data begin to flow. At present, no focused programs exist specifically for analysis of LSST solar system data}. Unlike NASA or DOE programs which typically come with a science team, funding, and key projects to be executed, Rubin Observatory is considered a {\em facility}, with no established funding for solar system science. This is in marked difference to some international partners who already have preparatory LSST science funding and are gearing up for first science. A real risk exists that -- having fully funded its M\$$600+$ construction -- the U.S. astronomy community may miss out on key LSST discoveries. {\bf The Decadal Survey should encourage NSF, NASA, and DOE to develop new Rubin-specific grant programs to ensure this facility fulfils its full potential as a solar system exploration machine.} Given that most LSST discoveries will occur during the first few years of the survey, preparedness and early investment will be crucial.

%support LSST Solar System science and develop synergies with complementary missions through existing and new grant programs. Areas in need of investment include, but are not limited to, research and analysis funding, computational and data management infrastructure, follow-up observation capabilities and management, and community coordination and work effort management (further details to be provided in a future white paper). Given that most LSST discoveries will occur during the first few years of the survey, early investments will be crucial.

%    \item
Upon the completion of the 10-year LSST survey, many important additional solar system science investigations will be possible with an extended or ``Phase 2'' survey.  
    As such, {\bf planetary science should be a high-priority consideration in planning the post-LSST future of Rubin Observatory.}
    %An extended survey could also make many additional discoveries of NEOs, PHAs, comets, and active asteroids that were not observable during the initial LSST survey.
    %(e.g., those that require longer time baselines or that would benefit from changes to the original survey design such as longer exposure times, cadence changes, or different filters) 
%\end{itemize}

%Draft findings
%\begin{itemize}
%    \item Extended solar system survey
%    \begin{itemize}
%        \item Additional discoveries: NEOs and SP comets that could not be observed in first 10-years, PHAs, new long-period comets
%        \item Statement about spectroscopy?
%        \item improved orbital characterization: Yarkovsky (needs longer baselines), YORP, cometary non-gravs, better orbits for distant objects?
%        \item Allows us to address our next priorities in the Science Roadmap.
%        \item Consider longer exposure times. (and maybe also shorter exposure times to look at near-planet environments; i.e., search for satellites?)
%        \item Are there proposed filters that align with our science?
%    \end{itemize}
%    \item Summarize our other WP goals, as we see them now (R\&A funding, computing etc.)
%    \item Make sure Planetary Science is a consideration of Rubin Observatory's ``Phase 2''.
%    \item Discovery of future mission targets (primary, flyby, and extended mission targets).
%\end{itemize}

\vspace{-17pt}

\bibliography{references.bib}

\end{document}